\documentstyle[prl,aps,epsbox]{revtex}
\begin{document}
\draft
\preprint{}
\twocolumn[\hsize\textwidth\columnwidth\hsize\csname@twocolumnfalse%
\endcsname 
\title{
Duality and Anholonomy
in Quantum Mechanics
of 1D
Contact Interactions 
}
\author{
Izumi Tsutsui${ }^{1}$, 
Tam\'{a}s F\"{u}l\"{o}p${ }^{2}$ 
and 
Taksu Cheon${ }^{1,3}$
}
\address{
${ }^1$
Theory Group, 
High Energy Accelerator Research Organization (KEK),
Tanashi, Tokyo 188-8501, Japan\\
${ }^2$
Institute for Theoretical Physics,
Roland E\"{o}tv\"{o}s University,
P\'{a}zm\'{a}ny 
P.s\'{e}t\'{a}ny 1/A, 
Budapest H-1117, 
Hungary\\
${ }^3$
Laboratory of Physics, Kochi University of Technology,
Tosa Yamada, Kochi 782-8502, Japan\\
}
\date{March 29, 2000}
%
\maketitle
\begin{abstract}
We study systems with parity invariant 
contact interactions in one dimension.
The model
analyzed is the simplest 
nontrivial one --- 
a quantum wire with a point defect ---
and yet is shown to
exhibit exotic phenomena,
such as strong vs weak coupling
duality and spiral anholonomy in the
spectral flow.  
The structure underlying these phenomena
is $SU(2)$, which arises as accidental symmetry
for a particular class of interactions.
\end{abstract}
%
]
\narrowtext

%
%
%
Throughout its century-long history, 
quantum mechanics has been mostly the
science of explanation for 
naturally existing
microscopic objects.
With the advent of nano-scale engineering,
however, we are witnessing the emergence of
quantum technology, in which 
a system of desired
specification is designed and 
manufactured \cite{EL00}.
It is commonly assumed
that
the operations of
nano-scale devices can be
understood, in essence, in
terms of the quantum mechanics of
elementary textbooks.
This by no means diminishes the 
significance of
the remarkable
technological achievements,
but one might still ask the question why
they do not admit the
richness of exotic phenomena, 
such as quantum anomaly, supersymmetry 
and duality, which grace
the frontier 
of high energy physics.
One might wonder
whether this is a limitation inherent 
to the low energy 
physics,
where the description by
simple quantum mechanics is sufficient,
without any real need for invoking
quantum field theory or string theory.

In this Letter, we wish to 
purge such pessimism by investigating
one of the simplest
quantum systems --- a
particle in one dimension 
subject to a contact interaction, 
an interaction that acts 
only at a single spatial point.  This is 
an idealized model of a quantum wire with a
single defect. 
It can be also regarded as a solvable limit
of a generic quantum system
in which the potential has an effective range
far shorter than the wavelength of 
the particle.
Contrary to the widely held view, the quantum
contact interaction in one dimension is
far from trivial\cite{AG88}.  
In addition to the well-known
Dirac's $\delta$-function potential 
which induces the
discontinuity in the spatial derivative of wave
functions, there exists another  
contact potential, called an `$\varepsilon$-function
potential', which brings about a
discontinuity in wave functions 
\cite{GK85,SE86a}.  
By combining $\delta$ and $\varepsilon$-function
potentials, one obtains a family of generalized 
one-dimensional contact interactions
described by a set of four parameters.
Several curious features of the 
generalized contact interaction have already been
pointed out \cite{AE94,EX96,CH98,CS99} over
the past few years.  We have seen, for instance, 
evidence of {\it duality} 
in the sense
of an isospectral mapping  between eigenstates of 
$\delta$ and $\varepsilon$-function 
potentials with reversed 
coupling strengths\cite{CS99}.  We also
know that there arises 
{\it spiral anholonomy} displaying the 
double spiral structure of the energy surface
\cite{CH98}.  (Contact interactions in two 
dimensions have been shown to possess 
scale anomaly \cite{JA95}.)
However,
these phenomena seem to be unrelated, and
their physical and mathematical origins 
have not previously been uncovered.

This is precisely the subject of our discussion 
in this Letter.
Specifically, we examine the
structure of 
parity invariant contact interactions, and
thereby uncover the general structure of
the duality and the spiral anholonomy
found earlier.  
In the course of our analysis,
we shall also find accidental $SU(2)$ symmetry
as well as supersymmetry 
that appears for a special
class of contact interactions.

%
%
%
%
\begin{figure}
\label{fig1}
\center\psbox[hscale=0.42,vscale=0.42]{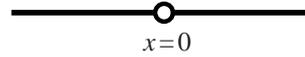}
\vspace{3mm}
\caption{
A particle moves freely on a line except at
the singular point, $x = 0$.
}
\end{figure}
We begin by considering
a quantum particle moving freely
on a line ${\bf R}$ except at a single 
point, $x = 0$, where a singular contact 
interaction
is at work.  This system is described 
by removing the singular point from the
line (Fig.1) and using the formal Hamiltonian, 
\begin{eqnarray} 
\label{F1} 
H = -{{\hbar^2}\over{2m}}{{d^2}\over{dx^2}}\ ,
\end{eqnarray}
defined on 
${\bf R}\setminus\{0\}$.
We look for
the most general condition
for allowable quantum mechanical motion.
%
To this end,
we follow 
the approach of Ref.
\cite{FT00} 
and define the two-component vectors,
\begin{eqnarray} 
\label{F2} 
\Phi =
  \left( {\matrix{{\varphi (0_+)}\cr
                  {\varphi (0_-)}\cr}
         } 
  \right),
\qquad 
\Phi' =
  \left( {\matrix{{ \varphi' (0_+)}\cr
                  {-\varphi' (0_-)}\cr}
         } 
  \right) .
\end{eqnarray}
from the values and derivatives of a wave function
$\varphi(x)$ at the left $x = 0_-$ 
and the right $x = 0_+$ of the missing point.
Due to the gap at $x = 0$, 
the wave function can acquire 
discontinuity there.  The allowed
discontinuity is dictated by 
quantum mechanical
probability conservation,
which requires the 
probability current 
$j(x) = - i\hbar(
(\varphi^*)'\varphi - \varphi^* \varphi' )/(2m)$
to be
continuous at $x = 0$. In terms of $\Phi$ 
and $\Phi'$, the requirement is expressed as
%
\begin{eqnarray} 
\label{F2a}
\Phi'^\dagger \Phi - \Phi^\dagger \Phi' = 0
\  ,
\end{eqnarray}
which is equivalent to
$|\Phi-i L_0 \Phi'|$  $=$ $|\Phi+i L_0 \Phi'|$
with $L_0$ being an arbitrary constant in the
unit of length.  
This means that, with a two-by-two unitary matrix
$U\in U(2)$, we have the relation,
\begin{eqnarray} 
\label{F3} 
(U-I)\Phi+iL_0(U+I)\Phi'=0
\ .
\end{eqnarray}
This shows that the entire family $\Omega$ of
contact interactions admitted in quantum mechanics
is given by the group $U(2)$.
Mathematically, we say that the domain
in which the Hamiltonian $H$ becomes self-adjoint
is parametrized by $U(2)$ --- there is
a one-to-one correspondence between
a physically distinct contact
interaction and a self-adjoint Hamiltonian.
A standard parametrization for $U$ is 
\begin{eqnarray} 
\label{F4} 
U=e^{i\xi }
  \left( {\matrix
            {{ \alpha_R+i\alpha_I}&{\beta_R+i\beta_I}\cr
             {-\beta_R+i\beta_I}&{\alpha_R-i\alpha_I}\cr}
         } 
  \right) 
\ ,
\end{eqnarray} 
where $\xi \in [0,\pi)$ is an angle and 
$\alpha_R$, $\alpha_I$, $\beta_R$ and $\beta_I$ are
four real numbers constrained by
\begin{eqnarray} 
\label{F5} 
\alpha_R^2+\alpha_I^2+\beta_R^2+\beta_I^2=1 .
\end{eqnarray}
%

For 
$\beta_R \beta_I \ne 0$
the condition (\ref{F3}) can be cast into the 
familiar
$U(1) \times SL(2,{\bf R})$ transition 
matrix form that
describes the gap effect at 
the singular point \cite{AG88}.
For $\beta_R \beta_I = 0$, on the other hand,
the condition (\ref{F3}) leads to a set of
separate
boundary conditions on the right and left sides of
the singular point \cite{AD98}.
The $U(2)$ form of
the boundary condition (\ref{F3}) 
provides a {\it global} description of the entire 
parameter space $\Omega$ in contrast to the 
conventional
$U(1) \times SL(2,{\bf R})$ description
which is local.

%
%
%
There are several
physically distinct subfamilies in the
parameter space $\Omega$ $\simeq U(2)$.
Here we concentrate on 
one 
which is perhaps physically most interesting,
that is, the subfamily consisting of 
parity invariant contact interactions.
The {\it parity} transformation is defined
by 
\begin{eqnarray} 
\label{P1}
{\cal P}: \, \varphi (x) 
\rightarrow  
({\cal P}\varphi)(x) := \varphi( - x),
\end{eqnarray}
under which 
the vectors in (\ref{F2}) transform as
\begin{eqnarray} 
\label{P2}
\Phi 
\buildrel {\cal P} \over \longrightarrow \sigma_1 \Phi\ , 
\qquad
\Phi' 
\buildrel {\cal P} \over \longrightarrow \sigma_1 \Phi'\ ,
\end{eqnarray}
where $\sigma_1$ is the Pauli 
matrix.  
Thus, parity
invariant boundary conditions arise if
we have
\begin{eqnarray} 
\label{P3}
\sigma_1 U \sigma_1 = U\ ,
\end{eqnarray}
which is satisfied 
for $\alpha_I=0$ 
and $\beta_R=0$ in (\ref{F4}).  
The solution can be most conveniently
parametrized as
\begin{eqnarray} 
\label{P4}
U = U(\theta_+, \theta_-)
  = e^{i\theta_+P_+}\, e^{i\theta_-P_-} \ ,
\end{eqnarray}
where the angle
parameters 
$\theta_\pm \in [0, 2\pi)$, 
and the projection operators $P_\pm$ 
are defined, respectively, by
\begin{eqnarray} 
\label{P5-P6}
\theta_\pm
&:=& \xi \pm \arctan{(\beta_I/\alpha_R)},
%
%
\\
P_\pm 
&:=& {1\over2}(1 \pm \sigma_1).
\end{eqnarray}
The operators $P_\pm$  fulfill the relations
$P_\pm^2 = P_\pm$ , $P_\pm P_\mp = 0$
and $P_+ + P_- = 1$.
Eq.~(\ref{P4}) shows that parity invariant 
contact interactions form a
subfamily $\Omega_P$ given by the torus of
two $U(1)$ groups,
\begin{eqnarray} 
\label{P7}
\Omega_P \simeq U(1) \times U(1) 
\subset \Omega \simeq U(2)\ .
\end{eqnarray}
We note that the decomposition 
$\Phi = P_+\Phi + P_-\Phi$ 
of any vector $\Phi$ 
corresponds to the decomposition
of the wave function 
into its
parity-symmetric and parity-antisymmetric parts, 
$\varphi(x) = \varphi_+(x) + \varphi_-(x)$ with
$\varphi_\pm(-x) = \pm \varphi_\pm(x)$.  
The parity projection splits 
Eq. (\ref{F3})
into two separate conditions for
even and odd parity eigenstates,
\begin{eqnarray} 
\label{P8}
\varphi_+(0_+)
\sin{\theta_+ \over 2}
+
\varphi'_+(0_+) L_0
\cos{\theta_+ \over 2} 
=0\ ,
\\ 
\nonumber
\varphi_-(0_+)
\sin{\theta_- \over 2} 
+
\varphi'_-(0_+) L_0
\cos{\theta_- \over 2}
=0\ .
\end{eqnarray}
A crucial point here is that
all physical properties of even (or odd) 
parity states are
now determined solely by 
the parameter $\theta_+$ (or $\theta_-$).

%
%
%
We take a closer look at the parity invariant
torus $\Omega_P \simeq S^1 \times S^1$.
Since the system splits into 
even and odd sectors, it is possible
to define the {\it coupling strength}
of the interactions independently in 
each sector.  Note, however, that 
there is an inherent asymmetry 
between the two sectors.  Namely, in the even 
sector, the system becomes free ({\it i.e.}, 
both $\varphi$ and $\varphi'$ 
are continuous at $x = 0$) when one has
$\varphi'_+(0_+)=0$, 
whereas in the odd sector 
the free system is achieved
with $\varphi_-(0_+) = 0$.
These occur at
$\theta_+ = 0$  
in the even sector and $\theta_- = \pi$
in the odd sector.
This observation 
leads us to the following definition of 
the strength 
$g_+$ ($g_-$) for the even (odd) sector,  
\begin{eqnarray} 
\label{D1}
g_+  &:=&  \tan(\theta_+/2) \ , 
\\ \nonumber
g_-  &:=&  \cot(\theta_-/2) \ .
\end{eqnarray}
We learn from (\ref{D1}) that there exist
contact interactions which act only on even
(or odd) parity sector
leaving the other sector free, and if we are 
to express
the interactions 
in terms of potentials, 
these are nothing but 
$\delta$ and $\varepsilon$-functions.
More explicitly, in the
$\theta_+$--$\,\theta_-$ plane (Fig.~2),
the $\delta$-function interactions
show up on the horizontal line
$\theta_- = \pi$, and the $\varepsilon$-function
interactions arise on the vertical line
$\theta_+ = 0$.
Their joint
$(\theta_+, \theta_-)$ $= (0, \pi)$ is the
point where both of the sectors become
free, {\it i.e.}, it represents 
the free system.

We now define the 
{\it half-reflection} transformation ${\cal R}$ by
\begin{eqnarray} 
\label{D2}
{\cal R}: \, \varphi (x) 
\rightarrow  
({\cal R}\varphi)(x) := [\Theta(x) -
\Theta(-x)]\varphi(x)\ ,
\end{eqnarray}
where $\Theta(x)$ is the Heaviside step function.
This causes the parity
interchange of states 
$\varphi_+$ and $\varphi_-$, which is implemented 
on vectors $\Phi$ and $\Phi'$ as
%
\begin{eqnarray} 
\label{D3}
\Phi 
\buildrel {\cal R} \over \longrightarrow \sigma_3
\Phi\ , 
\qquad
\Phi' 
\buildrel {\cal R} \over \longrightarrow \sigma_3
\Phi'\ .
\end{eqnarray}
Like the parity ${\cal P}$ transformation, 
${\cal R}$ induces a map
on the parameter space $\Omega \simeq U(2)$.
In particular, in the parity invariant subfamily
$\Omega_P$ we find
\begin{eqnarray} 
\label{D4}
U(\theta_+, \theta_-) 
\buildrel {\cal R} \over \longrightarrow 
\sigma_3\, U(\theta_+, \theta_-)\, \sigma_3
= U(\theta_-, \theta_+)\ ,
\end{eqnarray}
that is, the half-reflection transformation 
(\ref{D2})
induces the interchange 
$\theta_+ \leftrightarrow \theta_-$
yielding the mirror transformation with respect to 
the diagonal axis $\theta_+$ $= \theta_-$ 
in the $\theta_+$--$\,\theta_-$ plane.
It follows that the invariant
subfamily $\Omega_{SD}$ under the
half-reflection transformation 
is given by the set of points on the diagonal
line $\theta_+ = \theta_-$, 
which we call {\it self-dual},
which forms 
a $U(1)$ subgroup of $\Omega_P$,
namely,
\begin{eqnarray} 
\label{D5}
\Omega_{SD} \simeq U(1) \subset \Omega_P \simeq 
U(1) \times U(1)\ .
\end{eqnarray}

\begin{figure}
\label{fig2}
\center\psbox[hscale=0.42,vscale=0.42]{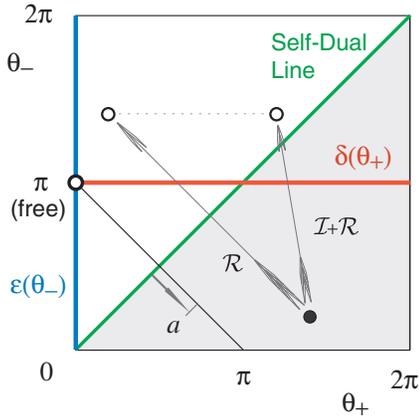}
\vspace{3mm}
\caption{(color)
The dissected torus $\Omega_P$.  Various
transformations 
are indicated by the arrows.
The horizontal red line and the vertical blue line 
represent the
$\delta$-function interactions and
the $\varepsilon$-function interactions,
respectively.
}
\end{figure}
If 
$\varphi$ is an energy eigenstate, 
$H\varphi = E \varphi$,
then the mapped state ${\cal R} \varphi$ 
is also an 
eigenstate with the same energy
$E$, but in general, of different 
connection condition.
In mathematical terms, the domain of $H$ specified 
by $U$ is changed under an isospectral 
transformation ${\cal R}$, 
and $H$ does not commute with ${\cal R}$
in general.
The physical consequence of the half-reflection 
transformation 
${\cal R}$ is now evident:
the spectra of even states and odd states are
interchanged leaving the whole spectrum unchanged.  
The point is that this is a 
{\it strong vs weak coupling
duality}, because under ${\cal R}$ 
the coupling strengths, 
$g_+$ and $g_-$ defined in (\ref{D1}), 
are interchanged and then inverted,
\begin{eqnarray} 
\label{D6}
g_+ \buildrel {\cal R} \over
\longrightarrow  {1 / g_-} ,         
\qquad             
g_- \buildrel {\cal R} \over
\longrightarrow 
             {1 / g_+} .
\end{eqnarray}
For illustration, take 
the point $(\theta_+,\theta_-) = (\pi, 0)$,
for instance.
Under ${\cal R}$ this is
dual to the free point $(0, \pi)$, and hence the
spectrum built on it is that of the free system
even though 
it is not free (since both of the
coupling strengths diverge there).
The situation should become clearer 
by inspecting Fig.~3, in which 
the momentum flow $k$, 
which is directly related to
the spectral flow by $E=(\hbar^2k^2)/(2m)$, is
plotted as a function of the distance from 
the self-dual line.  
%
\begin{figure}
\label{fig3}
\center\psbox[hscale=0.42,vscale=0.42]{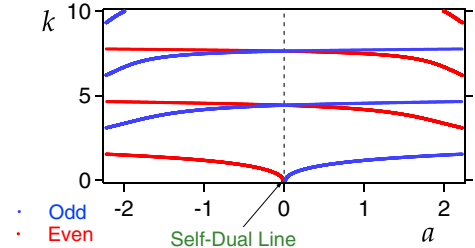}
\vspace{3mm}
\caption{(color)
Momentum flow $k$ along the line in Fig.~2
connecting
$(0, \pi)$ and $(\pi, 0)$ obtained under
the Dirichlet boundary conditions,
$\varphi_\pm(L)$
$=\varphi_\pm(-L)$
$=0$ for some $L$.
Two points with an equal distance from
the self-dual line 
in the opposite direction
are dual partners connected by ${\cal R}$.
}
\end{figure}

We define
yet another transformation by
\begin{eqnarray} 
\label{A1}
{\cal Q}: \, \varphi (x) 
\rightarrow  
({\cal Q}\varphi)(x) := 
 i[\Theta(-x) - \Theta(x)]\varphi(-x)\ .
\end{eqnarray}
This is just the combination 
${\cal Q}=-i{\cal R}{\cal P}$ and 
implements the transformation on vectors, 
\begin{eqnarray} 
\label{A2}
\Phi 
\buildrel {\cal Q} \over \longrightarrow \sigma_2
\Phi\ , 
\qquad
\Phi' 
\buildrel {\cal Q} \over \longrightarrow \sigma_2
\Phi'\ .
\end{eqnarray}
%
One then observes that
the transformations, ${\cal P}$, ${\cal Q}$ and
${\cal R}$, 
anti-commute each other,
and obey exactly the same relations
as those of the Pauli matrices:
\begin{eqnarray} 
\label{A3}
{\cal P}{\cal Q} 
= i{\cal R} , \ \ 
{\cal Q}{\cal R} 
&=& i{\cal P} , \ \  
{\cal R}{\cal P} 
= i{\cal Q} ,
\\ \nonumber
{\cal P}^2 ={\cal Q}^2
&=&{\cal R}^2= 1.
\end{eqnarray}
%
The existence of the $su(2)$ algebra
formed by 
$\{{\cal P}, {\cal Q}, {\cal R}\}$ 
suggests that, in the preceding analysis
(\ref{P1})-(\ref{D6}), we may exchange the role of
${\cal P}$ and ${\cal R}$,
or replace either of them by ${\cal Q}$.
For example, in the ${\cal R}$-invariant 
subfamily defined by $\sigma_3 U \sigma_3 = U$,
the generator
${\cal P}$ (or ${\cal Q}$) acts as the
operator  for isospectral interchange of states.

We now see why the double degeneracy
takes place along the self-dual
diagonal line $\theta_+$ $= \theta_-$.
On this line the generators
do not alter the domain of $H$, and hence
they commute with $H$.   
In other words, at self-dual points, 
all of the three
generators $\{{\cal P}, {\cal Q}, {\cal R}\}$
keep the boundary condition intact,
and hence the $SU(2)$ generated by 
them becomes {\it symmetry} there.  
One might also
wonder if there arises {\it supersymmetry} 
on the line in view of the fact that 
the degeneracy
occurs with symmetric and antisymmetric states.
Here we simply state the fact that, 
at the middle point
$(\pi, \pi)$, the system indeed becomes a
supersymmetric
Witten model \cite{JU96}, while
other self-dual points do not
admit a similar interpretation because the
supercharges fail to be  
self-adjoint there.

Discrete transformations 
may also be defined directly as a map
on the space $\Omega_P$ rather than being induced
by transformations on states.  For example, 
we consider 
\begin{eqnarray} 
\label{A4}
U(\theta_+, \theta_-) 
\buildrel {\cal I}_+ \over 
\longrightarrow 
e^{i\pi P_+}\,U(\theta_+, \theta_-) 
&=& U(\theta_+ \pm \pi, \theta_-)\ ,
\\ \nonumber
U(\theta_+, \theta_-) 
\buildrel {\cal I}_- \over 
\longrightarrow 
U(\theta_+, \theta_-) \, e^{i\pi P_-}
&=& U(\theta_+, \theta_- \pm \pi)\ ,
\end{eqnarray} 
which are the translations 
in the $\theta_+$--$\,\theta_-$
plane by a half-cycle (the sign $\pm \pi$ depends on
where $\theta_\pm$ lies).  
These are {\it coupling inversions} (with minus sign)
because their
effects on coupling strengths are
\begin{eqnarray} 
\label{A5} 
g_+ \buildrel {\cal I}_+ \over 
\longrightarrow -{1 / g_+} ,
\qquad
g_- \buildrel {\cal I}_- \over 
\longrightarrow -{1 / g_-} .
\end{eqnarray}
By combining ${\cal R}$
and ${\cal I}_\pm$ as
${\cal I}_+ {\cal R}$ $= {\cal R} {\cal I}_-$
or
${\cal I}_- {\cal R}$ $= {\cal R} {\cal I}_+$, 
one obtains the transformation that brings a
$\delta$-function interaction to an
$\varepsilon$-function interaction, and vise versa.
Note that only even spectra of $\delta$ and
odd spectra of $\varepsilon$ are identical
because of the presence 
of ${\cal I}_+$
and ${\cal I}_-$ which
change the spectra of the other sectors.
Thus we learn that the even-odd duality 
found earlier \cite{CS99} is a special
case of a wider class of duality found
in the whole $U(2)$ parameter space.


%
%
%
%
\begin{figure}
\label{fig4}
\center\psbox[hscale=0.42,vscale=0.42]{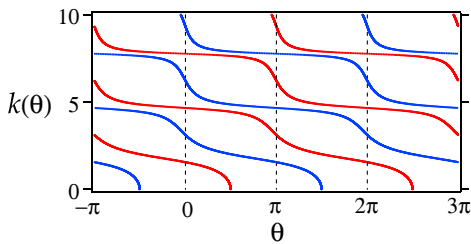}
\vspace{3mm}
\caption{(color)
The origin of the spiral anholonomy ---
the eigenvalues $k$ as a function of
the angle $\theta$ $= \theta_+$ $= \theta_- + \pi$.  
(The $\pi$ shift ensures the curve's passage 
through the free point $(\theta_+, \theta_-)$
= $(0, \pi)$.)
The cycle $\theta$ $\rightarrow$ $\theta + 2\pi$ 
leaves the spectra unchanged as a whole, 
but each level gets
shifted when one follows the
spectral flow.
}
\end{figure}
We now examine the origin of
the double spiral anholonomy in energy
levels found in Ref.\cite{CH98}. 
We look at the momentum eigenvalues $k$ as
functions of the parameters $\theta_\pm$.  
Under the
Dirichlet boundary condition
$\varphi_\pm(L)$
$=\varphi_\pm(-L)$
$=0$,  
the eigenvalues arise as solutions of
\begin{eqnarray} 
\label{S1}
k L_0\cot kL &=& \tan (\theta_+/2) \,
\ \ \ ({\rm even \ states}),
\\ \nonumber
k L_0\cot kL &=& \tan (\theta_-/2) \,
\ \ \ ({\rm odd \ states}).
\end{eqnarray}
Since any solution,
$k = k(\theta_+)$ or $k = k(\theta_-)$, 
of (\ref{S1}) is
a monotonously decreasing function as depicted in
Fig.~4, it is evident that each energy level acquires 
a spiral anholonomy as one completes a cycle along
the spectral curve in the
parameter space.  The double spiral 
arises when the cycle is completed simultaneously
in the two parameters $\theta_+$ and $\theta_-$, 
on account of the fact that even and
odd eigenstates arise alternately in the spectrum.
%

Finally, we mention that
an immediate extension of our study of quantum 
contact interactions is in the
analysis of junctions of more
than two lines \cite{EX96,KO99}.   In the case of a
junction of two lines, or the `x-junction', 
the problem
can also be regarded as contact interactions 
on a line for a particle with spin, pointing 
to the formulation 
of relativistic contact interactions \cite{GS86}.
In the light of the fact that some of them have
a potential application in designing
a device for quantum computation \cite{KO99}, 
their analysis along 
the line of the present work may be worth pursuing.
We end this paper with the restatement 
of our contention
that there is still nontrivial physics 
left to be discovered in the seemingly innocent 
settings of 
low-energy quantum mechanics.

\bigskip
T.C.~thanks 
Prof.~T.~Shigehara for helpful discussions. 
This work has been supported in part by 
the Grant-in-Aid for Scientific Research (C)
(No.~10640301 and No.~11640396) by 
the Japanese Ministry of
Education, Science, Sports and Culture.

\end{document}